# *Ultrafast jamming of electrons into an amorphous entangled state.*

Yaroslav Gerasimenko[2], Igor Vaskivskyi[2], Jan Ravnik[1,3], Jaka Vodeb[1,3], Viktor V. Kabanov[1] and Dragan Mihailovic[1,2,3]

*[1] Jozef Stefan Institute, Jamova 39, 1000 Ljubljana, Slovenia.*

*[2]CENN Nanocenter, Jamova 39, 1000 Ljubljana, Slovenia.*

*[3]Dept. of Physics, Faculty for Mathematics and Physics, Jadranska 19, University of Ljubljana, 1000 Ljubljana, Slovenia*

**New emergent states of matter in quantum systems may be created under non-equilibrium conditions if - through many body interactions - its constituents order on a timescale which is shorter than the time required for the system to reach thermal equilibrium. Conventionally non-equilibrium ordering is discussed in terms of symmetry breaking[1-5], nonthermal order-disorder [6-8], and more recently quenched topological transitions[9]. Here we report a fundamentally new and unusual metastable form of amorphous correlation-localized fermionic matter, which is formed in a new type of quantum transition at low temperature either by short pulse photoexcitation or by electrical charge injection in the transition metal dichalcogenide 1T-TaS$_2$. Scanning tunnelling microscopy (STM) reveals a pseudo-amorphous packing of localized electrons within the crystal lattice that is significantly denser than its hexagonally ordered low-temperature ground state, or any other ordered states of the system. Remarkably, the arrangement is not random, but displays a 'hyperuniform' spatial density distribution commonly encountered in classical jammed systems, showing no signs of aggregation or phase separation. Unexpectedly for a localized electron system, tunnelling spectroscopy and multi-STM-tip surface resistance measurements reveal that the overall state is gapless and conducting, which implies that localized and itinerant carriers are resonantly entangled. The amorphous localized electron subsystem can be understood theoretically to arise from strong correlations between polarons sparsely dispersed on a 2D hexagonal atomic lattice, while itinerant carriers act as a resonantly coupled reservoir distinct in momentum space. Apart from creating an entirely new may-body state of matter, the experiments reveal a path for the creation of unexpected emergent states that cannot be created or anticipated under thermodynamic equilibrium conditions.**

Materials with multiple competing interactions are ideal systems for searching new exotic states of matter forming under nonequilibrium conditions. In low-dimensional materials such as transition metal dichalcogenides, charge density waves (CDWs) [10] compete with Coulomb interactions, lattice strain and spin or orbital ordering, leading to breakup of such states under different circumstances and the formation of a variety of competing orders. The diversity of such states may be extended further under external influence. 1T-TaS$_2$ (Fig. 1a) is a well-studied dichalcogenide model system whose phase diagram is extraordinarily rich and includes metallic, incommensurate (IC), nearly commensurate (NC) and commensurate (C) CDW states, as well as superconductivity[11,12], and an unusual gapless quantum spin liquid state[13]. Its phase diagram is extended by photoexcitation, leading to an unusual metastable double vortex charge-ordered structure [9], while short-lived *transient* electronic ordering has been reported under different conditions[14]. Here we report on the creation of an exotic ultra-dense metastable ungapped *amorphous* electronic state in an otherwise perfectly ordered 1T-TaS$_2$ crystal, created by external laser pulses or charge injection, whose properties are without parallel in solid state fermionic systems.

An STM image of the C ground state (Fig. 1c) of 1T-TaS$_2$ at 4.2K from which we start, shows a CDW which can be viewed as a hexagonal packing of localized polarons, appearing as a triangular pattern of S atoms on the surface around the central Ta atom, enhanced by the chosen tip bias voltage ($V_t = $ 100 mV or -800mV). It is surrounded by 12 Ta atoms displaced towards it forming the shape of a hexagram (inset to Fig. 1c). After exposure of a $d \sim 100 \mu m$ diameter spot to a single 30fs, 800 nm laser pulse with fluence adjusted above a threshold of $F \simeq 3.5$ mJ/cm$^2$, the STM image reveals a dramatic change (Fig. 1d,e). The polarons now show an amorphous (A) arrangement, densely filling the entire surface. Remarkably, we see no long range density fluctuations, which is a signature of hyperuniformity[15]. A similar effect over an area $l^2 \sim 1 \times 1$ µm is observed when a $\sim 10$V pulse from an STM tip is passed through the sample. The amorphous structure is completely stable at 4.2 K during scanning. Unexpectedly, under sequential STM scanning conditions at 77 K the system becomes perceptibly *more* amorphous with time (See SI for Fourier transforms of consequtive STM scans), showing no tendency of relaxation towards the C ground state. The barrier for relaxation to the C state is thus clearly quite high, which is confirmed by the fact that the unperturbed state is stable for long periods (days) up to $200K$. However, 3V pulses from the STM tip (see SI) can cause local patches of C order to gradually appear. The fluence-dependence of spectral features of the collective mode measured by coherent phonon spectroscopy reveals that the transformation is not limited to the surface layer but is present up to a depth of at least 20 nm (see SI for details).

Optimizing STM contrast and tip bias allows us to identify the relative positions of localized electrons and compare their pair distribution functions (PDF) $g^{(2)}_{pol}(r)$ in the A state with the C ground state (Fig. 2b). Two things become apparent: (i) the A state has a much smaller nearest neighbor (NN) distance $d$ ($1 \pm 0.05$ nm) than the C state and (ii) there is a notable absence of any peaks at $r > 3$ nm which implies an absence of long-range correlations beyond 3 NNs. A similar PDF analysis for the atomic pattern $g_{at}^{(2)}(r)$ (Fig. 2c) reveals that the smaller NN positions found in polaronic PDFs (Fig. 2b) coincide with the 5$^{th}$ atomic neighbor, demonstrating that in the A state electrons predominately cause distortions on atomic sites which correspond to NN polaron-polaron distances. Clearly separated peaks at large distances (Fig. 2d) signify that the underlying lattice is only locally distorted without perturbation of long range lattice order.

Plotting the normalized radial dependence of the polaron density $\sigma$, $\sigma_R^2 = \left(\frac{\sigma}{R/D}\right)^2$ vs. $R/D$, where $D$ is the nearest-neighbor distance, we see a clear decay, which is a signature of a hyperunifrom distribution common in classical jammed systems[16]. A fit (Fig.2e) reveals a hyperuniformity exponent of $\beta = 0.60 \pm 0.03$. For comparison, random polaron packing would give $\beta = 0$, ($\sigma_R^2$ = constant). (See SI for detailed analysis). Remarkably, the density of localized polarons in the A state is ~ 23 % higher ($n_A = 0.98 \pm 0.02 \, nm^{-2}$) than the hexagonally packed C ground state ($n_C = 0.80 \pm 0.02 \, nm^{-2}$); a surprising fact, considering that the hexagonal packing of charges in the C state is the densest possible packing of classical spheres in 2D. Amorphous matter that is denser than the crystalline form is very rare in nature[17]. The remarkable stability (to > 200K) and absence of relaxation towards the C state implies that polarons are trapped in configurational space by the strong mutual correlations, exhibiting the kind of non-ergodicity that is expected for maximally correlated jammed states [15].

Investigating the electronic structure by tunneling spectroscopy reveals that the density of electronic states (DOS) is convincingly gapless for all polaron configurations (Fig. 3a) with a small V-shaped dip in the range ± 0.1 eV from $E_F$ which can be interpreted as a signature of incipient localization by some fraction of the carriers[18]. In contrast, the C state shows distinct sharp features commonly assigned to Hubbard bands[19-22] and zero conductance within the gap from −0.1V to +0.15V (Fig. 3a). To ascertain if the carriers in the A state are localized Fermi-glass-like or are metallic as the DOS suggests, we performed an in-situ three-STM tip switching experiment (see Fig. 1b for schematic). Two tips ~ 1 μm apart are used to apply a 30mA switching pulse and measure the resistance $R_{2tip}$ before and after switching. The third tip in between them is used to confirm the local A polaron structure after switching. The result is unambiguous: the resistance drops from $R_{2tip} \simeq 8k\Omega \rightarrow 340\Omega$ upon C→A switching (see SI for more details.) Thus, both vertical tunneling and the surface current between two tips shows that the A state becomes more conducting in spite of the electrons being localized. We conclude that the system is in a mixed quantum state composed of spatially localized polarons entangled with itinerant carriers. This distinguishes it from conventional Fermi and Coulomb glasses, and most other complex states of matter.

Considering the transition mechanism from the excited state to the A state, the observed non-ergodic dynamics suggests it is formed through many body localization of a significant fraction of the available electrons through mutual interactions. The electronic band structure in the undistorted state of 1T-TaS$_2$ (appropriate for a laser-melted CDW) has a single Ta band in the vicinity of the Fermi level (Fig. 3d) which exhibits an unusual duality: in the $M-K$, $\Gamma-M$ and $\Gamma-A$ (interlayer) directions the band crossing $E_F$ is predicted to be very dispersive with fast itinerant electrons, while along the $\Gamma-Y$ direction it is virtually dispersionless over a large range of momenta and just above the Fermi level [23]. The carriers in this band have an unusually large effective mass over a large range of momenta. After photoexcitation or injection through the STM tip, the energetic carriers rapidly lose energy through mutual scattering and phonon emission, while increasing in density as a result of avalanche multiplication[24]. At some point in time the carriers which occupy the dispersionless $\Gamma-Y$ band, the massive carriers start to localise. If we consider the Coulomb repulsion as an analogue of hard-sphere interaction of classical systems, above a certain critical density all available space is filled by the localizing carriers and the electrons become immobilized as a result of mutual correlations, following the classical density-driven jamming transition scenario. After the process is over, the rest of the carriers remain itinerant, occupying the dispersive bands crossing the Fermi level. The localized states are spread in momentum space over $\Delta k \simeq 2\pi/d$, i.e. over a large part of the zone, overlapping with the itinerant bands in k-space. However, having the same energy and occupying the same region of k-space as the localized particles, the two necessarily couple and become entangled.

Considering the localization dynamics on a regular 2D lattice[25,26], the strong tendency toward spin-charge separation and coexistence of localized charge order with spin-liquid-like fluctuations [13] reinforce theoretical indications that coupling of fermions to slow spin-fluctuations may help mediate the localization process resulting in non-ergodic behaviour[27,28]. Similarly, lattice vibrations are also expected to aid disordered localization, a notion supported by the experimental evidence for frozen lattice distortions in Fig. 2e. However, all such model discussions so far[27,29] lead to gapped behaviour, which is inconsistent with the measured DOS and itinerant electron transport. We conclude that the A state – with correlation-localized and itinerant carriers is beyond current single closed-system many body localisation theories, and requires a multiband treatment in which the localized electron states are resonantly entangled with itinerant electrons, to understand the experimentally observed itinerant electron character (Fig. 3d)

The stability of the correlation-localized electron subsystem, subject to Coulomb repulsion and lattice strain can be investigated with the Hamiltonian[30] $H = \sum_{i,j} V(i,j)(n_i - \bar{v})(n_j - \bar{v})$. Here $V(i,j)$ is a screened Coulomb potential, $n_i = 1$ if the site occupied and 0 otherwise, and $\bar{v}$ is the average polaron density. The configurational minimum within this model found by Monte-Carlo simulations of $H$ on a hexagonal mesh, using the experimental density $n_A$ as a fixed input parameter compares remarkably well with experiments in Fig. 4a and b respectively. (See SI for calculation details). Experimentally, our observation that relaxation proceeds in the direction of greater disorder (see SI) supports the implied notion that the A state is a free energy minimum. The model also predicts a hyperuniform distribution of polarons, the PDFs (Fig. 4f) showing quite good agreement with the STM data (Fig. 2d) and a comparable hyperuniformity exponent $\beta_{theory} = 0.87 \pm 0.04$ (Fig. 4 g). Screened Coulomb interactions are thus apparently sufficient for a qualitative understanding of the hyperuniform polaron distribution even without considering spin or lattice disorder. But since transport is limited to diffusion within such models[30-32], the metal-like transport requires consideration of the mixed state.

The effective charge $q^*$ of each polaron is shown by a Voronoi cell construction shown in Figs. 4 a, b for the experimental data and the model respectively. It reveals a tiling of irregular polygons with no translational or rotational symmetry. Generally $q^* \neq e$ so each electron is shared between different numbers of Ta atoms. Without additional doping, the C ground state structure has one electron per unit cell with 13 Ta atoms. Other electron-rich ordered structures with different electron number densities may exist in this structure with $\nu = \frac{1}{4}, \frac{1}{7}, \frac{1}{9}, \frac{1}{12}, \frac{1}{13}$ ... electrons per 1T-TaS$_2$ elementary unit cell whose primitive cells are shown in Fig. 4 e. The observed average experimental density in the A state is $\nu = 0.094e$. The state may thus be thought of as an electron-rich tessellation with tiles that correspond to the different charge ordered states, with an average between $\frac{1}{9}$ and $\frac{1}{12}$. The itinerant $\Gamma - M$ and $\Gamma - K$ bands then act as a charge reservoir to which the charged localized states are coupled. By analyzing the inter-polaron distances, we can assign corresponding primitive cells to each polaron (Figs. 4 c, d), which beautifully shows that the state shows no sign of phase separation or aggregation into any kind of regular structure.

## *Acknowledgments.*

We wish to thank Tomaz Prosen, Janez Bonča, Peter Prelovšek, Rok Žitko, Tomaž Mertelj, Serguei Brazovskii, Vladimir Dobrosavljevic and Petr Karpov for useful discussions, Jernej Mravlje for LDA calculations, Maksim Litskevich for STM measurements and Petra Sutar for synthesis and characterization of the samples. Funding from ERC-2012-ADG_20120216 "Trajectory" is acknowledged.

## *Author contributions*

IV made the original discovery, DM and YG lead the project, wrote the paper and performed analysis. YG, JR and IV performed the measurements, JV and VVK performed theoretical calculations. All authors contributed to the supplementary information.

*Figure captions.*

Figure 1. **Structure, experiment and basic observation by STM of initial and final states of polarons after excitation by a single laser pulse.** a) Schematic view of two layers of 1T-TaS$_2$ crystal. b) Low-temperature UHV STM configuration with multiple tips (3 are shown) combined with the ultrafast laser beam. c) C ground state exhibiting a regular lattice of polarons, shown schematically in the left insert. The right insert shows a (raw) atomic resolution STM image of polarons at 0.1 V tip bias emphasizing the S atoms. d) The A state formed after a single 30 fs pulse at 4.2K shows an amorphous structure. e) An atomic resolution image of the A-state shows localized polarons superimposed on the atomic lattice.

Figure 2. **Distances in between polarons and atoms.** a) A point pattern extracted from the experimental STM images of polarons in Fig. 1d. b) Polaron PDFs in the A and C states in blue and orange respectively derived from a). The peak assignments for the C-state in terms of the nearest neighbours are shown in c). d) A point pattern of atoms extracted from Fig. 1e. e) Atomic PDFs of the A and C states shown by blue and green lines respectively. The orange line is a simulation for the C state. The PDF in the A state is smeared at 1 nm and 1.15 nm, corresponding to the nearest-neighbour polaron distance (see blue dashed lines). f) Polaron density fluctuations in the A-state for the data within the radius shown in a). The blue line is a power law fit to $\left(\frac{\sigma}{R/D}\right)^2$ giving a hyperuniform value of $\beta = 0.6 \pm 0.03$. (see SI for details of fit).

Figure 3. **Density of electronic states (DOS) measured by STS** a) Comparison of the C and A states (the latter is averaged over 25 different points on the surface shown in the topographic image panel (c)). Red dots indicate where spectroscopy curves were measured. While the C-state has clear gap at the Fermi level, the A-state has finite DOS and a small linear V-shaped pseudogap-like feature centred at $E_F$. The vertical dashed lines indicate the C state band edges which appear to be partly preserved in the A-state. Vertical grey regions indicate the position of the states commonly ascribed lower and upper Hubbard bands. b) Individual DOS curves measured at different points (white squares in c)), illustrating the robustness of the DOS and the pseudogap. d) a schematic in-plane high temperature band structure of 1T-TaS$_2$ (after [23]). The shaded regions correspond to itinerant and localisation-prone states respectively. The insert indicates the directions in the Brillouin zone.

Figure 4. **Tiling patterns formed by polarons in the A state shown by two types of mapping (expt. and modelling).** a) and b) reveal the effective polaron charge for each Voronoi cell which varies from 1/7 to 1/16 $e^-$ per atomic unit cell. c) and d) show tiling based on the characteristic distances between the nearest neighbors in the pattern. While tile area varies, its diagonal corresponds to one of the five characteristic superstructures on the triangular lattice shown in the primitive cell e). Remarkably, we see almost no aggregation in the experimental data (e). f) Comparison of the experimental and Monte-Carlo simulations of polaronic PDFs, shown in blue and orange respectively. Dashed vertical lines show the positions of the peaks corresponding to specific superstructures (e). g) Dependence of a normalized variance on distance for the point pattern obtained in Monte-Carlo simulations shows hyperuniform behavior with the decay rate of $\beta_{theory} = 0.87 \pm 0.04$, compared with the measured value $\beta_{expt.} = 0.6 \pm 0.03$ shown in Fig. 2.

1. Nasu, K., Ping, H. & Mizouchi, H. Photoinduced structural phase transitions and their dynamics. *J Phys-Condens Mat* **13,** R693–R721 (2001).
2. Zurek, W. Cosmological experiments in superfluid helium? *Nature* **317,** 505 (1985).
3. Yusupov, R. *et al.* Coherent dynamics of macroscopic electronic order through a symmetry breaking transition. *Nat Phys* **6,** 681–684 (2010).
4. Schmitt, F. *et al.* Transient electronic structure and melting of a charge density wave in TbTe3. *Science* **321,** 1649–1652 (2008).
5. Madan, I. *et al.* Evidence for carrier localization in the pseudogap state of cuprate superconductors from coherent quench experiments. *Nat Comms* **6,** 6958 (2015).
6. Iwai, S. *et al.* Ultrafast optical switching to a metallic state by photoinduced Mott transition in a halogen-bridged nickel-chain compound. *Phys. Rev. Lett.* **91,** 057401 (2003).
7. Koshihara, S., Tokura, Y., Mitani, T., Saito, G. & Koda, T. Photoinduced valence instability in the organic molecular compound tetrathiafulvalene-p-chloranil (TTF-CA). *Phys. Rev. B* **42,** 6853–6856 (1990).
8. Cavalleri, A. *et al.* Femtosecond Structural Dynamics in $VO_2$ during an Ultrafast Solid-Solid Phase Transition. *Phys Rev Lett* **87,** 237401 (2001).
9. Gerasimenko, Y. A., Vaskivskyi, I. & Mihailović, D. Long range electronic order in a metastable state created by ultrafast topological transformation. *arXiv 1704.08149v1* (2017).
10. Monceau, P. Electronic crystals: an experimental overview. *Adv Phys* **61,** 325–581 (2012).
11. Rossnagel, K. On the origin of charge-density waves in select layered transition-metal dichalcogenides. *J Phys-Condens Mat* **23,** 213001 (2011).
12. Sipos, B., Berger, H., Forro, L., Tutis, E. & Kusmartseva, A. F. From Mott state to superconductivity in 1T-TaS2. *Nature Materials* **7,** 960–965 (2008).
13. Klanjsek, M. *et al.* A high-temperature quantum spin liquid with polaron spins. *Nat Phys* **13,** 1130–1134 (2017).
14. Han, T.-R. T. *et al.* Exploration of metastability and hidden phases in correlated electron crystals visualized by femtosecond optical doping and electron crystallography. *Science* **1,** e1400173–e1400173 (2015).
15. Torquato, S. Hyperuniformity and its generalizations. *Physical Review E* **94,** 022122 (2016).
16. Torquato, S. & Stillinger, F. H. Local density fluctuations, hyperuniformity, and order metrics. *Phys Rev E Stat Nonlin Soft Matter Phys* **68,** 411131–4111325 (2003).
17. Mishima, O., Calvert, L. D. & Whalley, E. 'Melting ice' I at 77 K and 10 kbar: a new method of making amorphous solids. *Nature* **310,** 393–395 (1984).
18. Shklovskii, B. I. & Efros, A. L. *Electronic properties of doped semiconductors*. (Springer Series in Solid-State Sciences 45, 1984).
19. Fazekas, P. & Tosatti, E. Charge Carrier Localization in Pure and Doped 1T-TaS2. *Physica B & C* **99,** 183–187 (1980).

20. Hellmann, S. *et al.* Time-domain classification of charge-density-wave insulators. *Nat Comms* **3,** 1069– (2012).
21. Cho, D. *et al.* Nanoscale manipulation of the Mott insulating state coupled to charge order in 1T-TaS2. *Nat Comms* **7,** 10453 (2016).
22. Ma, L. *et al.* A metallic mosaic phase and the origin of Mott-insulating state in 1T-TaS2. *Nat Comms* **7,** 1–8 (2016).
23. Reshak, A. H. & Auluck, S. Full-potential calculations of the electronic and optical properties for 1T and 2H phases of TaS2 and TaSe2. *Physica B: Physics of Condensed Matter* **358,** 158–165 (2005).
24. Kabanov, V. V. & Alexandrov, A. S. Electron relaxation in metals: Theory and exact analytical solutions. *Phys Rev B* **78,** 174514 (2008).
25. De Roeck, W. & Imbrie, J. Z. Many-Body Localization: Stability and Instability. *arXiv.org* **math-ph,** 20160422 (2017).
26. Mahmoudian, S., Rademaker, L., Ralko, A., Fratini, S. & Dobrosavljević, V. Glassy Dynamics in Geometrically Frustrated Coulomb Liquids without Disorder. *PRL* **115,** 333 (2015).
27. Smith, A., Knolle, J., Kovrizhin, D. L. & Moessner, R. Disorder-Free Localization. *PRL* **118,** 201 (2017).
28. Smith, A., Knolle, J., Moessner, R. & Kovrizhin, D. L. Absence of Ergodicity without Quenched Disorder: From Quantum Disentangled Liquids to Many-Body Localization. *PRL* **119,** 201 (2017).
29. Nandkishore, R. & Huse, D. A. Many-body localization and thermalization in quantum statistical mechanics. *Annual Review of Condensed Matter Physics* **6,** 15–38 (2015).
30. Karpov, P. & Brazovskii, S. Multi-domain fragmentation and walls' globules in a 2D lattice of charged particles with applications to pump- and pulse induced hidden states in 1T-TaS$_2$. *arXiv 1709.01912v1* (2017).
31. Miranda, J. *et al.* Polaron and bipolaron transport in a charge segregated state of a doped strongly correlated two-dimensional semiconductor. *PRB* **83,** 125308 (2011).
32. Rademaker, L., Pramudya, Y., Zaanen, J. & Dobrosavljević, V. Influence of long-range interactions on charge ordering phenomena on a square lattice. *Physical Review E* **88,** 175 (2013).

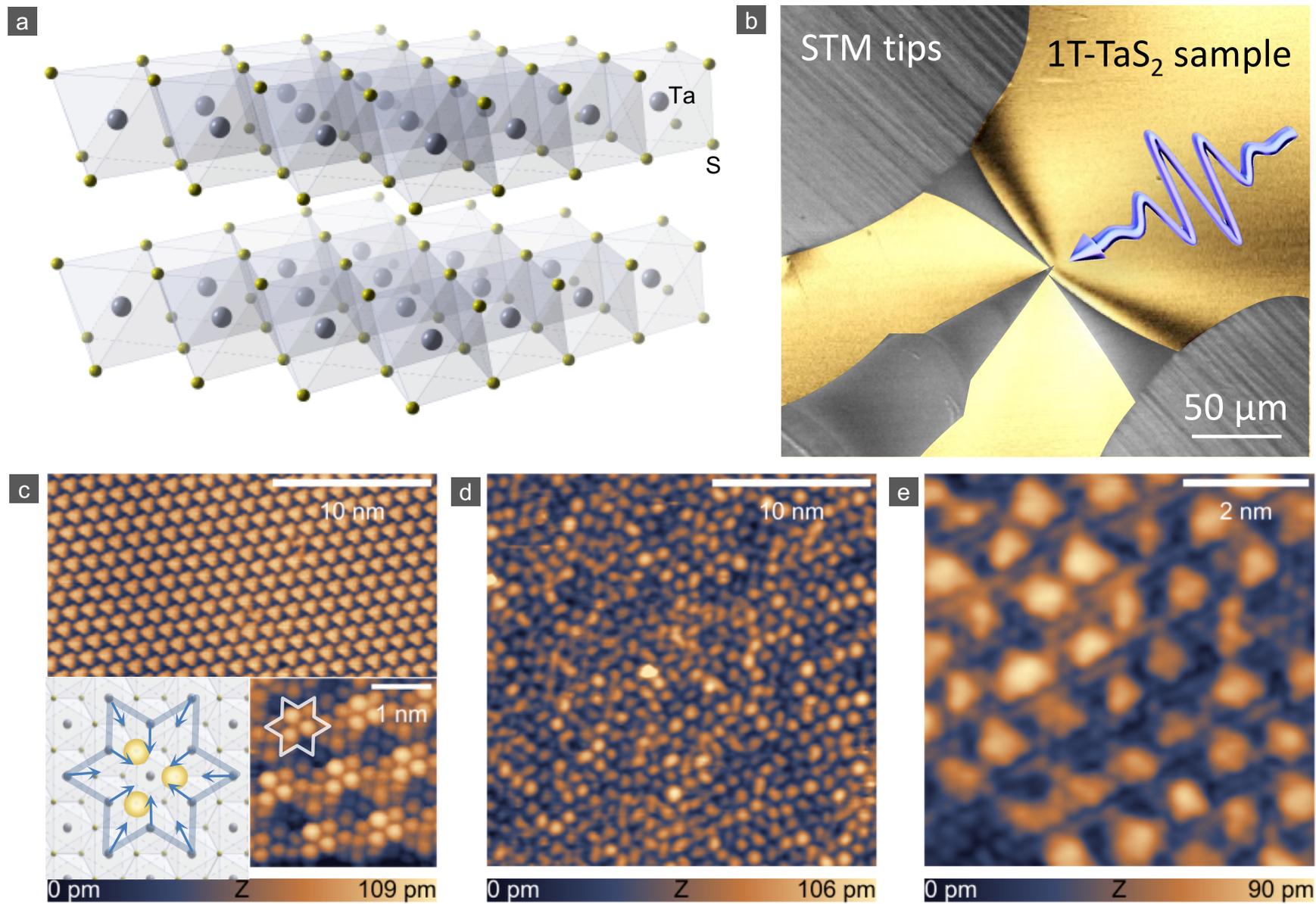

Fig. 1

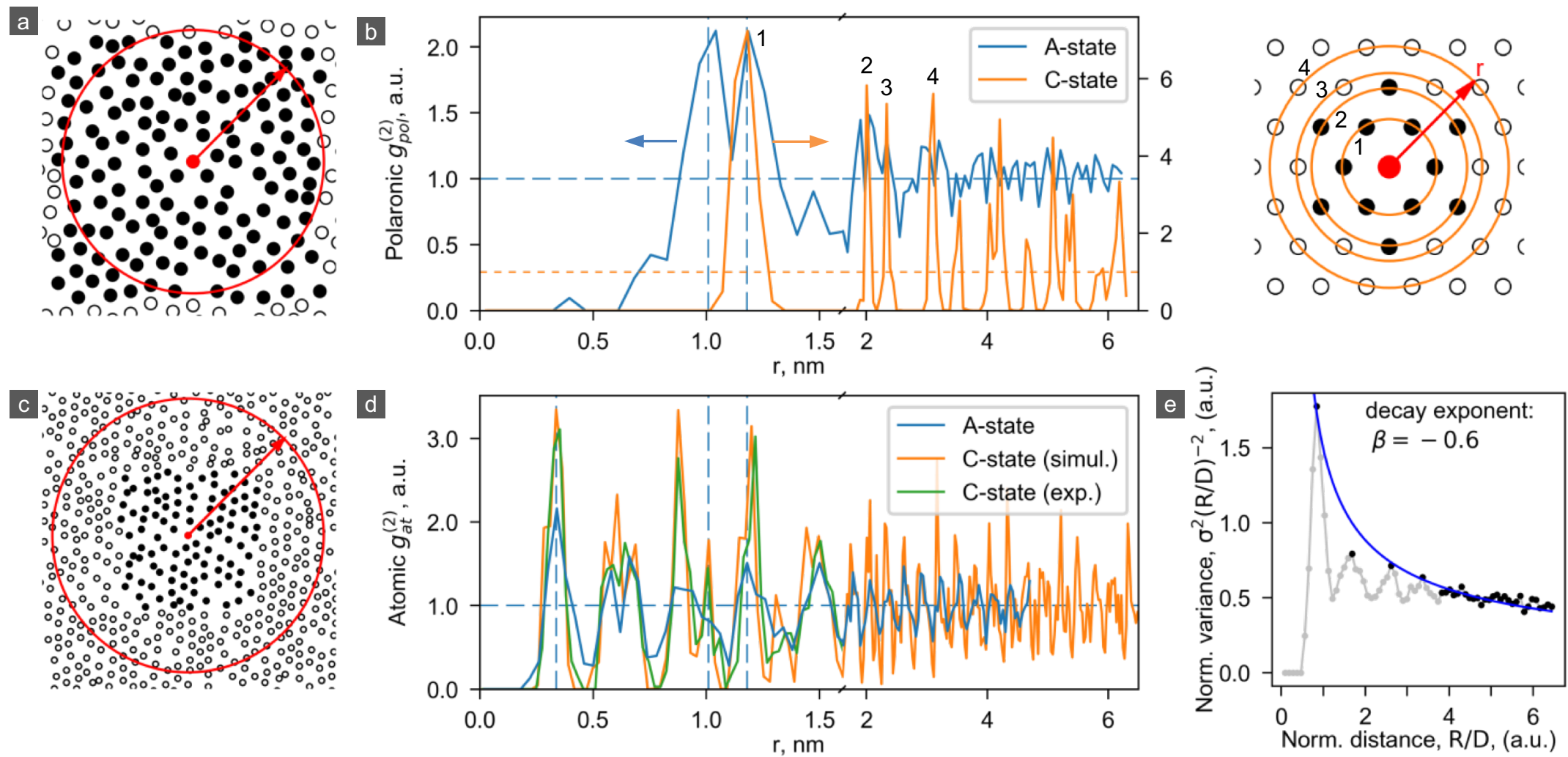

Fig. 2

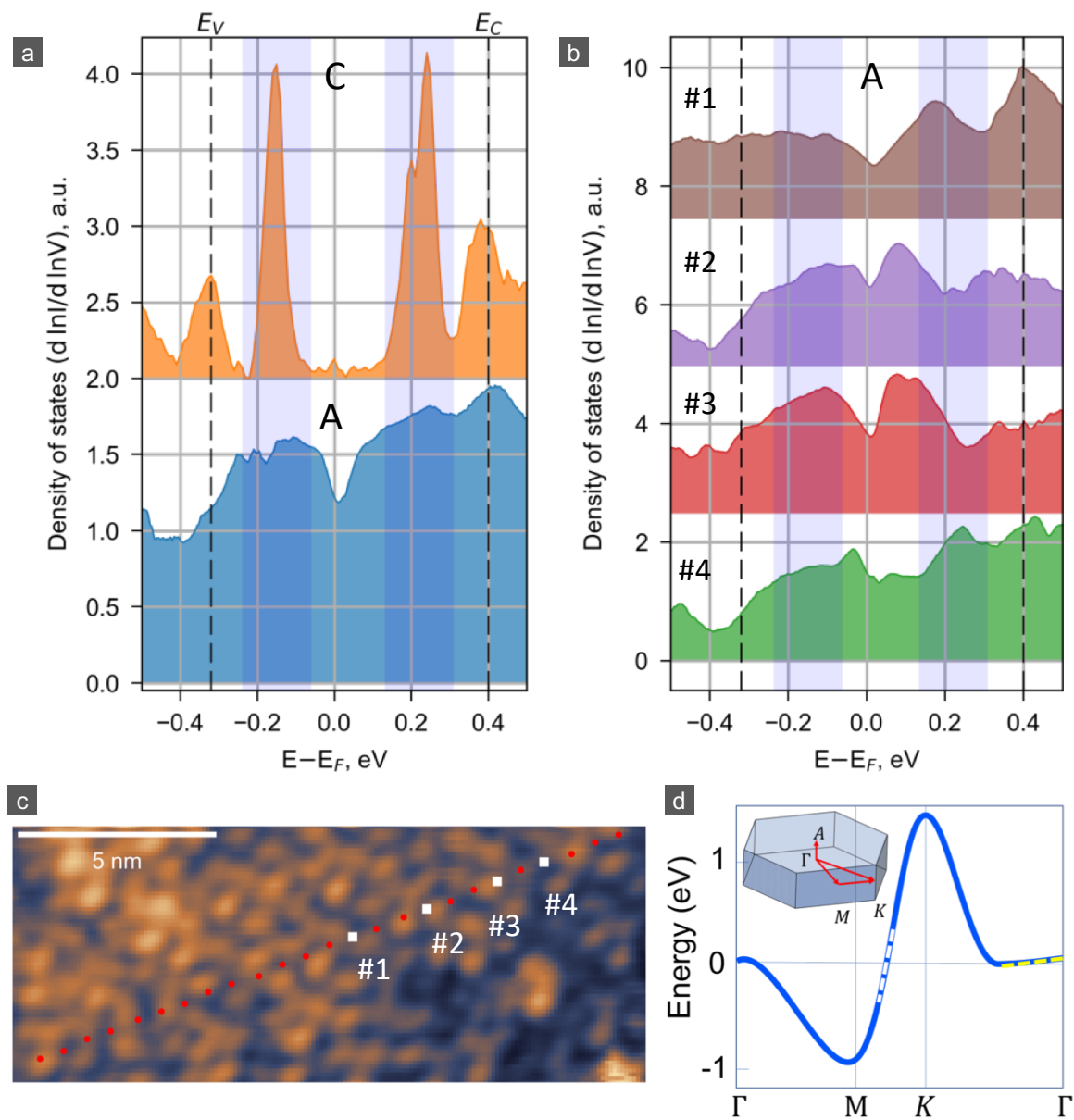

Fig. 3

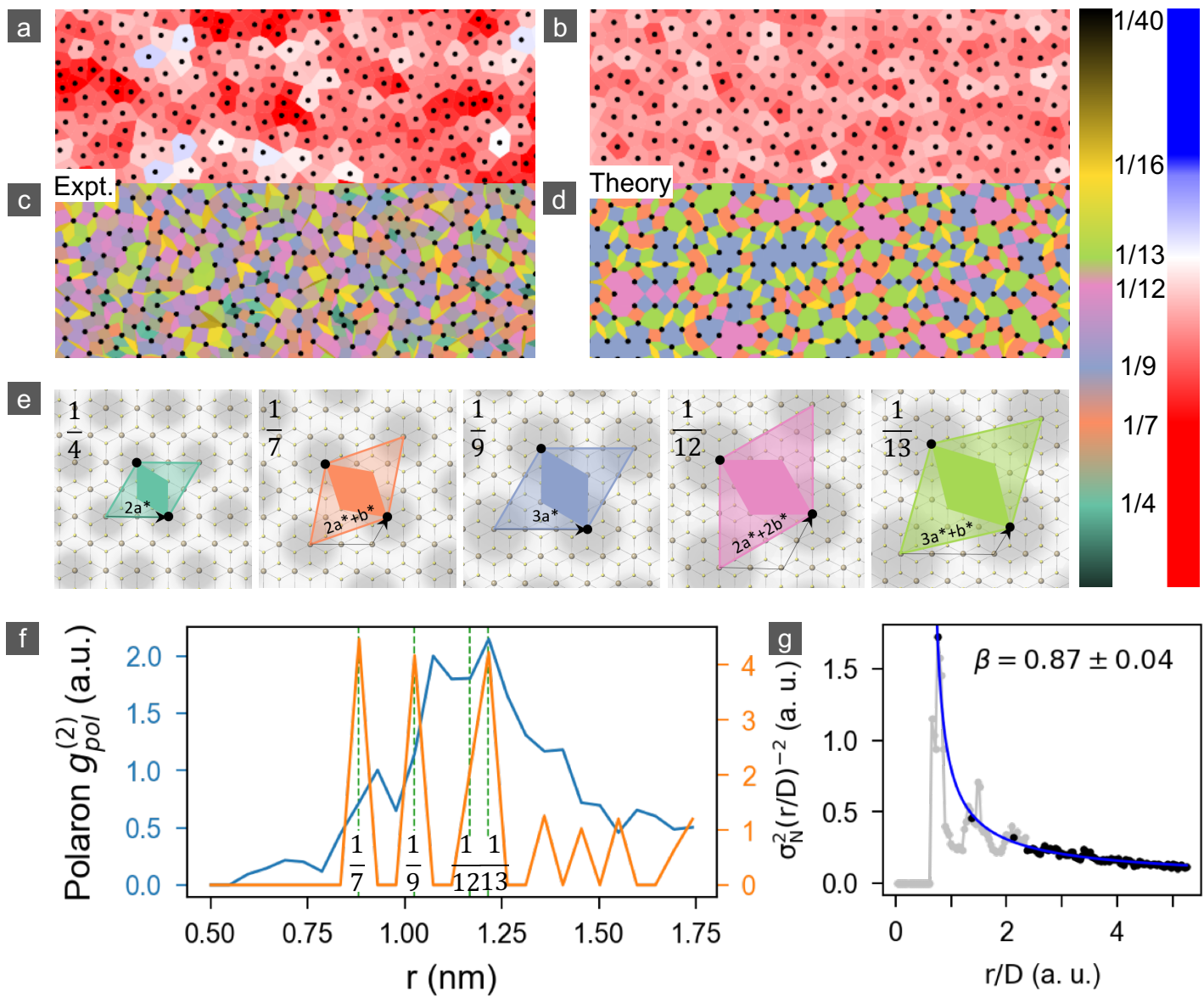

Fig. 4